\documentclass{elsart}
\usepackage{graphicx}% Include figure files
\usepackage{amssymb}
\usepackage{bm}% bold math
\newcommand{\pargp}{\hspace*{0.5in}}
\begin{document}
\begin{frontmatter}
%%%\height{210mm}
%%%\width{297mm}

%%%%%%%%%%%%%%%%%%%%%%%%%%%%%%%%%
\title{A Biologically Inspired Ratchet Model of Two Coupled Brownian Motors}
\author{Debasis Dan and A. M. Jayannavar}
\ead{dan@iopb.res.in, jayan@iopb.res.in}
\address{Institute of Physics, Sachivalaya
Marg, Bhubaneswar 751005, India}
\author{Gautam I. Menon}
\ead{menon@imsc.ernet.in}
\address{The Institute of Mathematical Sciences,\\ C.I.T. Campus,
Taramani, Chennai 600 113, India} 
\begin{keyword}
ratchet, molecular motor, Langevin equation.
\PACS 87.16.Nn, 05.40.-a
\end{keyword}
%%%%%%%%%%%%%%%%%%%%%%%%%%%%%%%%%%%%%%%%%
\begin{abstract}
A ratchet model for coupled Brownian motors,
inspired by the motion of individual two-headed
molecular motors on cytoskeletal filaments,
is proposed. Such motors are modeled as two
elastically coupled Brownian particles, each of
which moves in a flashing ratchet potential. The
ratchet potentials felt by the individual
particles are anti-correlated, modeling the
successive binding and unbinding of the two motor
heads to the filament. We obtain, via Langevin
simulations, steady-state currents as functions
of noise strength, the equilibrium separation of
the particles and the rate of switching between
potential states. We observe an enhanced current
due to coupling, noise induced stability and
phase-locked behaviour in the deterministic
regime. A qualitative understanding of these
features is provided.
\end{abstract}
\end{frontmatter}
\section{Introduction}
Motor proteins (kinesins, dyneins and myosins)
are versatile biomolecules which shuttle cargo
encapsulated in vesicles to different parts
of the cell. Such transport occurs {\it via}
the cytoskeleton, a cell-spanning polymeric
network of microtubules, actin filaments and
intermediate filaments. Motor proteins are also
important ingredients of the mechanisms of muscle
contraction and cell division. The quest for
physical principles that enable such tiny
molecular machines
to function efficiently in a highly
Brownian regime is a subject of ongoing interest.\cite{ajdari}  \\
%%%
\pargp Molecular motors are typically dimeric\cite{alberts}.
Conventional kinesins, for example, consist of
two identical proteins, each of which has a motor
domain (the head) and a cargo binding domain.
A stalk intervenes between the head and the
cargo binding domain. The two stalks coil around
each other to produce the dimer.  The two heads
walk unidirectionally along a micro-tubule,
a long and fairly stiff polymer comprised of
an asymmetric repeating unit, in 8 nm steps.
Each such step is coupled to the
hydrolysis of a molecule of ATP. 
Typically, such kinesins can walk several hundred 
steps before detaching from the micro-tubule track. \\
%%%
\pargp Simple models for molecular motors
idealize them as Brownian motors\cite{ajdari,menon}.  
A Brownian
motor is a point particle with an overdamped
equation of motion, which exhibits directed
motion when subject to thermal and athermal
noise\cite{ajdari,rev}.  The thermal noise is assumed to be
Gaussian and satisfies a fluctuation-dissipation
relation.  For molecular motors, the athermal
noise reflects the non-equilibrium driving
of the motor through the irreversible hydrolysis 
of ATP.\\
%%%
\pargp The interaction of the motor with the molecular
track is modeled in terms of an asymmetric
periodic potential felt by the motor\cite{ajdari}. This
potential switches between two or more states.
The energy release upon ATP hydrolysis is
believed to alter the coupling of the motor
protein to the substrate. It is thus modeled
by a switch in the potential state seen by
the Brownian particle. \\
%%%%
\pargp The fact that the motor molecule
is a two-headed object is ignored
in these simple models (however, see
Refs. \cite{mielke,ajdari2}).  Experimentally,
two structural elements of the junction of the
two heads appear to be crucial to rapid
motility --- the neck and the neck linker. The
neck linker lies just beyond the catalytic
core of the head, while the neck leads from the
neck linker to the coiled stalk.  The coupling
of the two heads {\it via} the neck region
ensures that an elastic interaction operates
between the heads.  The enhanced processivity
and rate of movement in {\it N. Crassa}\/
conventional kinesins has been attributed, in
part, to the enhanced flexibility of its neck
region\cite{kali,song}. \\
%%%
\pargp Models of motor activity also indicate that the 
binding of one head as a consequence of ATP hydrolysis
correlates closely with the loosening of the
other head upon the subsequent release of the
inorganic phosphate\cite{howard}. This suggests that a more
physically realistic Brownian motor model of the working of a
two-headed molecular motor would involve (a) the incorporation of
an elastic interaction between motor heads and 
(b) switching between
potential states which are fully anti-correlated
between heads, so that one head is loosely bound
when the other is tightly bound and {\it vice
versa}. \\
%%%
\pargp To develop these ideas further, we
propose and study a simple model of a two-headed
molecular motor interacting with a thermal bath
at temperature $T$. Our model system consists
of two elastically coupled overdamped Brownian
particles moving in periodic potentials $V_1(x)$
and $V_2(x)$. These potentials alternate in
time; we will assume this
switching to be time periodic with period
$\tau$. We take $V_{1}(x)$ to be a periodic
asymmetric potential and $V_{2}(x)$ to be a
flat potential.  The potentials felt by the two
particles are out of phase {\it i.e.} when one
of them moves under the influence of potential
$V_{1}(x)$, the other only sees the force due
to the elastic linkage with the other particle.\\
%%%
\pargp Our model has some similarities, and yet some
important differences, with others proposed earlier. 
Unlike the models proposed by Klumpp {\it et. al.}\cite{mielke}, 
and by Ajdari\cite{ajdari2}, our model has no multiplicative 
{\em noise} terms, although the time-periodic variation of the
potential does enter multiplicatively in our equations of
motion.  Also, in our model, the potentials ~$V_1$ or $V_2$~ 
felt by one head are (anti-)correlated with the potentials felt
by the other, not uncorrelated as in earlier work.\\
%%%
\pargp We have studied our model numerically {\it via}
Langevin simulations in the weak and strong
coupling regimes.  
Our model exhibits a range of unusual
behaviour, including an enhanced current due to
coupling in some parameter range (this has been noted
earlier in a different but related model \cite{mielke}), drift in
the deterministic regime (with phase-locked behaviour \cite{gitt}),
and noise induced stability \cite{mielke2}. Subsequent sections of
this paper describe these features.\\
%%%
\pargp In the next
section we propose the system of coupled Langevin
equations to describe the model. In Sec.~
\ref{det_motion} we present results for our model in the
deterministic regime, varying  the
coupling constant, the equilibrium separation of the
two particles and the frequency of potential
fluctuation. In Sec.~\ref{noise_motion} we discuss
the behaviour of the current in the presence of
thermal noise.  Our conclusions are contained in
Sec.~\ref{concl}.
%%%
\section{\label{model}Model}
\pargp We consider two overdamped Brownian particles, coupled 
{\it via} an elastic interaction,
moving in a two state flashing ratchet potential which switches between
$V_{1}(x) \mbox{ and } 0$ periodically. The particles are in contact
with a thermal bath at temperature $T$.  Let $x_{1}(t)$ and
$x_{2}(t)$ be the coordinates of the two particles. They are
coupled through a
spring of spring constant $k$ and equilibrium length $a$. The
Langevin equations which governs our system  are
%%%
%%\begin{subequations}
%% \label{langvn}
\begin{eqnarray}
\label{langvn}
 \gamma \dot{x_{1}} &=& -z_{1}(t)V'(x_{1})+k\big((x_{1}-x_{2})-a\big)+\xi_{1}(t) \\
 \gamma \dot{x_{2}} &=& -z_{2}(t)V'(x_{2})-k\big((x_{1}-x_{2})-a\big)+\xi_{2}(t)
\end{eqnarray}
%%\end{subequations}
%%%
where $\xi_{i}(t)$ are Gaussian random forces with zero mean and a
correlation given
by $<\xi_{i}(t)\xi_{j}(t')> = 2D\delta_{ij}\delta(t-t')$. Here, $D =
\frac{k_{B}\gamma T}{m}$. The $z_{i}$'s
are periodic functions with period $\tau$, given by 
$z_{1}(t)=1, z_{2}(t)=0  \mbox{ for } 0 \leq t < \frac{\tau}{2};
z_{1}(t)=0, z_{2}(t)=1  \mbox{ for } \frac{\tau}{2} \leq t <
\tau$. The asymmetric periodic potential with unit period is  
$V(x) = -\frac{U}{2\pi} (\sin(2\pi x) + 0.25\sin(4\pi x))$.
We work with dimensionless units wherever possible.\\ 
\pargp We solve these coupled equations 
numerically by Huen's method \cite{huen} and
calculate the current
in the asymptotic regime. The current $j$ is given by
$j = \big< \frac{x(t)-x(t_{0})}{t-t_{0}} \big>$, where $<..>$ denotes
the ensemble average and $t_{0}$ the asymptotic time. We take
$t_{0}=50 \tau, t=4000 \tau$ and average over $3000$ ensembles.
In the limiting case of rigid rod like coupling (i.e.,
$x_{1}-x_{2}=a$), we solve
the associated Fokker-Planck equation $\frac{\partial P(x,t)}{\partial t} = -\frac{\partial J(x,t)}{\partial x}$ ~\cite{risken}
  \begin{equation}
    \label{FPE}
    \frac{\partial P(x,t)}{\partial t} =
                      -\frac{\partial}{\partial
      x}\big( V'(x)+\sqrt{D}\frac{\partial}{\partial x}\big )P(x,t), 
   \end{equation}
for $n\tau < t \leq (n+1/2)\tau $,
  \begin{equation}
    \frac{\partial P(x,t)}{\partial t} = 
  	-\frac{\partial}{\partial
      x}\big( V'(x+a)+\sqrt{D}\frac{\partial}{\partial x}\big )P(x,t),
  \end{equation}
for $(n+1/2)\tau < t \leq (n+1)\tau$. 
Here $x$ is the position of one of the particle, the position of
the other particle being fixed at $x+a$ and $P(x,t)$ is the probability
density.  The average current 
%%%\begin{equation}
 $ j = \lim_{t \rightarrow \infty} \frac{1}{\tau} \int_{t}^{t+\tau} dt
    \int_{0}^{1} J(x,t) dx . $
%%%\end{equation}
We have obtained $j$ by numerically solving the Fokker-Planck equation.

\section{Results \lowercase{and} Discussions}
\subsection{\label{det_motion}Deterministic Current ($D = 0$)}
\pargp  A single particle moving in a flashing potential
such as the one described above, exhibits no directional
motion in the {\em absence} of thermal fluctuations.
However, such
motion can be induced in a system of elastically
coupled particles using the compressibility
of the spring ~\cite{ajdari2,mielke}. 
If the equilibrium separation of the two particles is smaller
than the length of the smaller arm of the potential $V(x)$
($L_{min}=0.38$) or larger than the longer arm
($L_{max}=0.62$), then
the phase space of the system is a closed orbit. No directed
current is possible for any value of $\tau$ or $k$.
%%%
%%%
%%%%%%%%%%%%%%%%%%%%%%%%%%%%%%
\begin{figure}[h]
\includegraphics[scale=0.75]{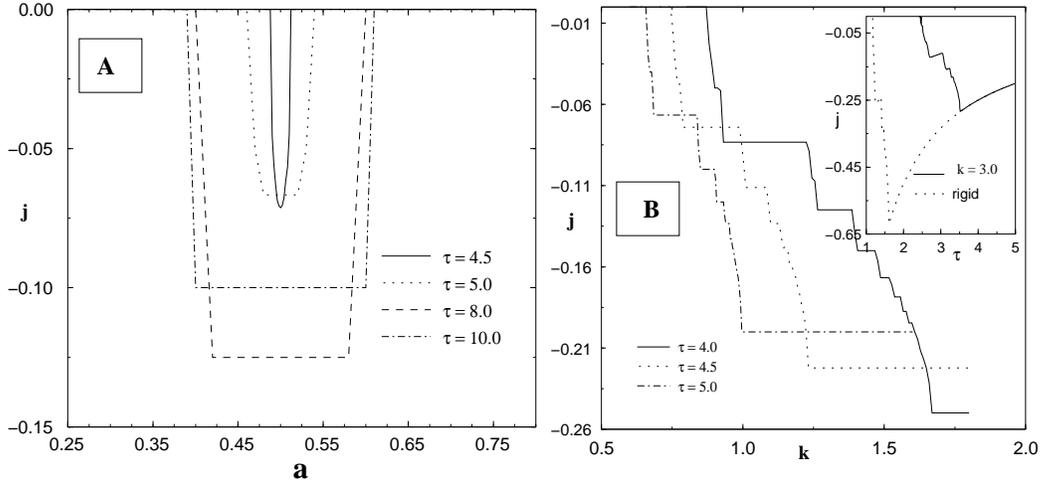}% Here is how to import EPS art
\caption{\label{j0-a}[A] Current $j \mbox{ \textit{vs}  equilibrium
    distance }a$ 
 at $k=0.8$; [B] current $j \mbox{ \textit{vs} coupling constant }k$ for 
 $a=0.5$. The inset shows $j \mbox{ \textit{vs} }\tau$ for same 
 parameter value.}   
\end{figure}
%%%%%%%%%%%%%%%%%%%%%%%%%%%%%
%%%%%%
%%%%%%
The
window $L_{max}-L_{min}$, for which a directed current is obtained
depends on the value of $\tau$ (Fig.(1A)).
For smaller values of $\tau$, the length of the
window for nonzero current is
small, as the particle has less time to traverse to the basin of
attraction of the next well.  In Fig.(1B) we  plot the current
%%%
%%\begin{figure}
%%\includegraphics[scale=0.5]{j_k.eps}% Here is how to import EPS art
%%\caption{\label{j0-k}}
%%\end{figure}
%%%
$j$ as function of $k$ for $a=0.5$, for different values of $\tau$.
Observe that the current exhibits a phase locked behaviour, with step
size becoming larger for smaller $\tau$ as observed previously in many different
systems. Such phase locking arises in deterministic systems due to the
interplay between nonlinearity and external drive. \\
\pargp For large $k$, the current saturates to a constant value, which
depends on $\tau$. At these values of $k$, the particles
traverse exactly one spatial period in one time period; since our
system is overdamped, no multiple hops can occur. The
inset shows the variation of $j
\mbox{ with } \tau$ for $k=3.0 \mbox{ and } k=\infty$ (rigid rod
case). The value of $\tau$ for which current is maximum decreases with
increasing $k$.
%%%%
%%%%
\subsection{\label{noise_motion} Effect \lowercase{of} Noise}
\pargp Many of the attributes of the deterministic
limit, such as the phase locked behaviour
of Fig. (1B), are sensitive to
noise. In the presence of thermal
fluctuations the noise-induced current exhibits 
many interesting features. There are three distinct regimes in the
parameter space where qualitatively different features are
obtained: (i) large $\tau$ and $L_{min} < a < L_{max}$, (ii) large
$\tau$ and $a < L_{min}$ and (iii) the small $\tau$ or 
nonadiabatic regime, which we do not discuss here.\\
%%%% 
\pargp We first study case (i), examining the effect
of noise in a parameter range where there is
current in the deterministic limit. In the large $\tau, a=0.5$ regime
the current $j$ increases  
initially with $k$ for all values of $D$ as shown in
fig.(2A). On further increasing $k$ (not shown) the
current saturates after a minimum.
%%
%%%
\begin{figure}[h]
\includegraphics[scale=0.75]{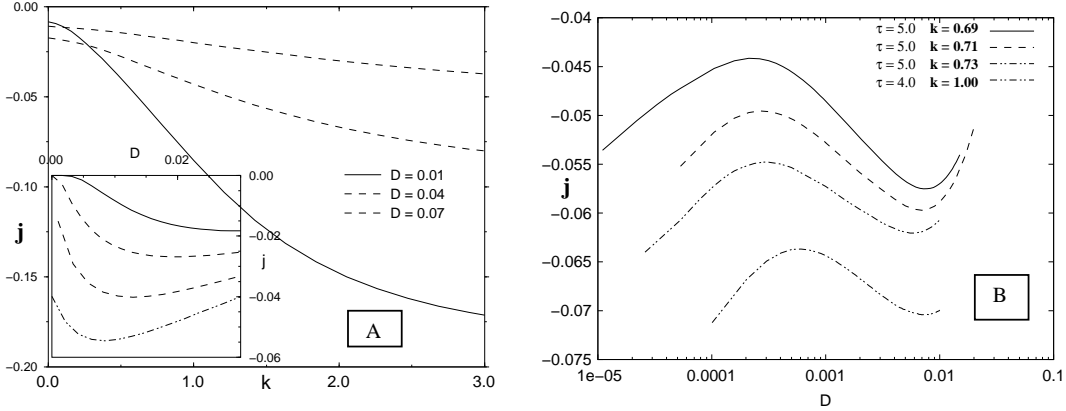}% Here is how to import EPS art
\caption{[A] Current $j$ \textit{vs} $k$ for $a = 0.5$. 
 The inset shows the plot of  $j$ \textit{vs} $D$ for $k =
  0.0, k= 0.3, k = 0.5 \mbox{ and } k = 0.67$.( From top to bottom
  curve.) [B] The peak in the $j \mbox{ \textit{vs} } D$ ( corresponding 
 to minimum current) emphasizes the noise induced stability phenomena. Here 
 $a = 0.5$. 
  \label{t5_a.5}}
\end{figure}
%%%
%%%%
The inset shows the variation of $j$ with $D$
for different values of $k$, keeping the  other
parameters the same. The case $k=0$ is the single particle
limit. The location of the peak in the $j-D$ curve is determined 
by the time taken to
diffuse to the basin of attraction of the next well ($j_{max}$~
at ~$D \approx L_{min}^2/\tau=0.16/5=0.032$). For
non-zero coupling, when one particle slides down the
potential the other is dragged along, thereby enhancing the
current. However, on further increasing the noise strength, the
potential landscape becomes irrelevant, random hopping in
both directions dominates and the absolute value of the current
tends to zero, accounting for the minimum seen. At
$k=\infty$, the current decreases monotonically with increasing 
$D$. \\
%%%
\pargp Though the increase of current with $D$ for low coupling strength or
monotonous decrease at higher $k$ is the generic feature in this
regime of operation, for certain values of $k$ \big(values for which
there is phase locking in the
deterministic regime, see Fig.~(1B)\big) the current shows highly
nonintuitive behaviour.
%%%
%\begin{figure}
%\includegraphics{nis}% Here is how to import EPS art
%\caption{\label{nis}}
%\end{figure}
As shown in Fig. (2B), the absolute current
initially decreases with increasing $D$, reaches a minima and then
again increases with $D$. This feature is reminiscent of the noise induced
stability (NIS) of unstable states ~\cite{mielke2}, seen previously in
different models, {\it i.e.} noise localizes the particle in its basin of
attraction. This leads to a decrease in the value of the current 
with noise.  In the high temperature regime, the current should go to
zero, explaining, qualitatively, the two extremas seen in the 
current as a function of noise. \\
%%
%%%
\begin{figure}[h]
\begin{center}
\includegraphics[scale=0.4]{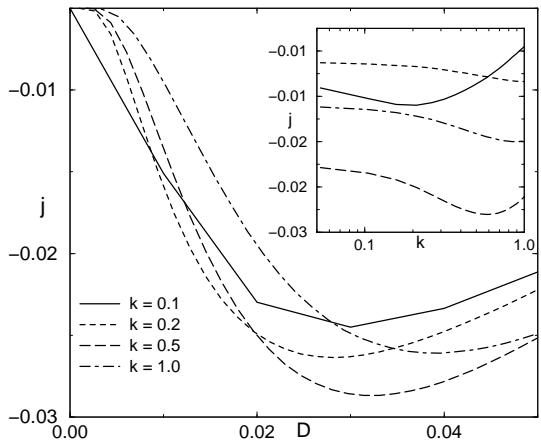}% Here is how to import EPS art
\caption{$j \mbox{ \textit{vs} } D$ for different values of $k$, $a = 0.2$. 
 The inset shows the variation of $j \mbox{ \textit{vs} } k$ in
 log scale, for $D = 0.01,~ 
 0.04, ~0.07 \mbox{and} ~0.1$ from top to bottom. 
\label{t5_a.2}}
\end{center}
\end{figure}
%%%%%
\pargp We now investigate the effect of noise 
in the parameter range where there is no current in the deterministic
limit, our case (ii). We work with $a = 0.2$.
Fig.  (\ref{t5_a.2}) shows $j$ \textit{vs} $D$ for various values 
of $k$. Unlike the
previous case where the temperature at which the current peaks decreased
with increasing $k$, here this temperature first decreases and 
then {\em increases} with increasing $k$. \\
\pargp The current here arises as a consequence of both the ``flashing
ratchet mechanism'' and the cooperative ``pulling effect''. For low
coupling, the contribution of coupling at a temperature at which the
current arising out of the ``flashing'' mechanism is maximum, 
is negligible. Hence the net
current peaks at a lower value of $T$ compared to the $k = 0$ case. For
higher $k$, where the current due to coupling dominates, a different
scenario obtains. As the stiffness of the spring is larger, a 
larger fluctuation is required for the particle to transit between 
wells. The larger the $k$, the
stronger the amplitude of the fluctuation required. If, however,
the temperature is too high,
significant motion is 
induced in the opposite direction and the current reaches a
maximum. Depending on the parameter regime, the coupling constant may
or may not enhance the current compared to the single particle
current. This can be clearly seen from Fig.~\ref{t5_a.2} and the
inset. This behaviour again differs from the case where there is
a finite current in the deterministic limit.\\
\pargp In the nonadiabatic
regime obtained for low values of $\tau$, even for extremely small values of $k$, 
currents may either enhance {\em or} decrease relative to the single particle
current. This differs from our earlier observation in the adiabatic
limit where an enhanced current is obtained for arbitrary small values
of $k$, for any $D$. These results will be reported elsewhere.
%%%
%
\section{\label{concl}Conclusions}
\pargp Our simple model of a two-headed motor comprised
of two coupled Brownian particles in
a flashing potential, exhibits unidirectional
current. The coupling between particles induces
an additional mechanism by which a current may be
obtained ({\it via} a pulling effect), which is
absent in the single particle case\cite{mielke}. The behaviour
of the current depends sensitively on system
parameters and exhibits several novel features,
such as an enhanced current due to coupling
compared to the single particle case, in a regime
which we identify.  Outside this regime, coupling
may or may not lead to a current enhancement.\\
\pargp As a function of noise strength, the current can
exhibit noise induced stability in a restricted
parameter range. Outside this range, noise
initially enhances current and the absolute
value of current shows a single maxima. In
the deterministic case, we have shown the
existence of a window for the two particle
separation distance within which there is a
deterministic current.  This current shows a
phase locked behaviour with coupling constant.
Unlike previous studies on related models, we
have not observed current reversals as a function
of system parameters.\\
\pargp What relevance, if any, might these results have
to real motors? While generically, weakening the
coupling of the two heads (particles) appears
to {\em reduce} the current, we do see specific
regimes in which such a reduction leads to an
{\em enhancement} in the current. As discussed
in the Introduction, the increased motility of
some Kinesins has been attributed to a
larger flexibility of the neck linker.  However,
whether such simplified models can ever reflect
the underlying behaviour of vastly more complex
biological systems is unclear.\\

\end{document}